# On the invariant thermal Proca – Klein – Gordon equation

M. Pelc

Physics Institute, Maria Curie – Sklodowska University, Lublin, Poland
e-mail: magdaanna@o2.pl

Abstract

In this paper we discuss the invariant thermal Proca - Klein - Gordon equation (PKG). We argue that for the thermal PKG equation the absolute velocity is equal $v = alpha*c$, where *alpha* is the fine stucture constant for electromagnetic interaction.

Key words: thermal equation, fine stucture constant.

1. Introduction

In the seminal book: "Special relativity theory", professor Andrzej Szymacha [1] formulated the STR with only one basic assumption: of the democracy of all inertial reference frames. In that formulation the Lorentz transformation has the form (in 1D case):

$$t = \frac{1}{\sqrt{1-\kappa^2 V^2}}\left(\kappa V x' + t'\right), \tag{1.1}$$

$$x = \frac{1}{\sqrt{1-\kappa^2 V^2}}\left(x' + V t'\right). \tag{1.2}$$

In the above formulae (1.1) and (1.2) $\kappa^2$ is the invariant constant which is independent of speed $V$. Only by comparison to the experimental data one obtains $\kappa^2 = 10^{-17}\ \mathrm{s^2/m^2} \approx \frac{1}{c^2}$, where $c$ is the light velocity in vacuum. However it must be stressed out that for different values of $\kappa$ which is independent of $V$ the Lorentz transformations (1.1) and (1.2) are valid prescription for $(x,t) \rightarrow (x',t')$.

In our paper we put forward the study of the previously formulated Proca – Klein – Gordon thermal equation [1]. It will be shown that the P – K – G equation can be rewritten in fully invariant form with $\kappa=(\alpha c)^{-1}$, where $\alpha$ is the fine structure constant.

2. The Proca – Klein – Gordon thermal equation

In the monograph [2] the relativistic thermal transport equation was formulated

$$\frac{1}{v^2}\frac{\partial^2 u}{\partial t^2} - \frac{\partial^2 u}{\partial x^2} + q(x,t)u(x,t) = G(x,t). \tag{2.1}$$

where

$$u(x,t) = T(x,t)e^{t/2\tau} \tag{2.2}$$

and $T(x,t)$ is the temperature of the system.

In Eq.(2.1) $\tau$ is the relaxation time for subquantum phenomena

$$\tau = \frac{\hbar}{m\alpha^2 c^2}, \tag{2.3}$$

where $m$ is the mass of the heat carriers, $\alpha$ is the coupling constant for electromagnetic interaction, $\alpha = \frac{e^2}{\hbar c}$ and $q$ is defined as:

$$q = \frac{2Vm}{\hbar^2} - \left(\frac{m\upsilon}{2\hbar}\right)^2. \tag{2.4}$$

In Eq.(2.4) $V$ is the external potential. Equation (2.1) is the thermal analog of the Proca – Klein – Gordon equation and can be described as:

$$\left(\bar{\Box}^2 + q\right)u(x,t) = G(x,y,z,t), \tag{2.5}$$

where d'Alembertian $\bar{\Box}^2$ is equal to (in 1D)

$$\bar{\Box}^2 = \frac{1}{\alpha^2 c^2}\frac{\partial^2}{\partial t^2} - \frac{\partial^2}{\partial x^2}. \tag{2.6}$$

The mass dependent term $q$ has the form:

$$q = \frac{\bar{m}^2 \alpha^2 c^2}{\hbar^2}, \tag{2.7}$$

where

$$\bar{m}^2 = \left(\frac{2Vm}{\alpha^2 c^2} - \frac{m^2}{4}\right). \tag{2.8}$$

Considering formulae (2.7) and (2.8) Eq.(2.5) can be written as

$$\left(\bar{\Box}^2 + \frac{\bar{m}^2 \alpha^2 c^2}{\hbar^2}\right)u(x,t) = G(x,t). \tag{2.9}$$

It is interesting to observe that the term $q = \frac{\bar{m}^2 \alpha^2 c^2}{\hbar^2}$ has the meaning the reciprocity "Bohr radius" for the heat transport on the quantum scale, for

$$\frac{\bar{m}^2 \alpha^2 c^2}{\hbar^2} = \frac{\bar{m}^2 e^4}{\hbar^4} = \frac{1}{R_B^2}. \tag{2.10}$$

With formula (2.10) equation (2.9) has the form

$$\left(\bar{\Box}^2 + \frac{1}{R_B^2}\right)u(x,t) = G(x,t) \tag{2.11}$$

or in compact form

$$\left(\partial_\mu \partial^\mu + \frac{1}{R_B^2}\right)u(x,t) = G(x,t), \tag{2.12}$$

where

$$\partial_\mu = \left(\frac{\partial}{\partial(v,t)}, -\vec{\nabla}\right). \tag{2.13}$$

Equation (2.12) is the thermal Proca equation, with $R_B \neq \infty$, i.e. $\bar{m} \neq 0$. For $\bar{m} = 0$, $R_B = \infty$, Eq.(2.14) is the thermal Klein – Gordon equation:

$$\partial_\mu \partial^\mu u(x,t) = G(x,t).$$

3. Conclusions

In this paper we present the invariant form of the thermal Proca – Klein – Gordon equation. It must be stressed that for the thermal Proca – Klein – Gordon equation the absolute velocity is the product $\alpha c = \frac{e^2}{\hbar}$ and not $c$.

[1] A. Szymacha, *Special Relativity Theory* (in Polish), Alfa 1985.
[2] M. Kozlowski, J. Marciak – Kozlowska, *Thermal Processes Using Attosecond Laser Pulses*, Springer 2006.